\let\ifarxiv=\iftrue     
\pdfoutput=1

\ifarxiv

\documentclass[12pt,a4paper]{article}
\usepackage[a4paper,text={450pt,650pt},centering]{geometry}

\fi

\ifarxiv\else

\documentclass[11pt,a4paper]{article}
\usepackage{mathptmx}
\usepackage[a4paper,text={130mm,198mm}]{geometry}

\fi

\usepackage{amssymb}
\usepackage{amsmath}

\usepackage{amsmath,amssymb}
\usepackage[bookmarks=true,hyperfigures=true]{hyperref}
\usepackage{graphicx}
\usepackage[nosort]{cite}
\usepackage[bulletsep]{collref}

\let\oldbfseries=\bfseries
\let\oldmdseries=\mdseries
\let\oldnormalfont=\normalfont
\renewcommand{\bfseries}{\oldbfseries\boldmath}
\renewcommand{\mdseries}{\oldmdseries\unboldmath}
\renewcommand{\normalfont}{\oldnormalfont\unboldmath}

\allowdisplaybreaks[3]

\numberwithin{equation}{section}

\usepackage[font=small,labelfont=bf,width=0.85\textwidth]{caption}

\providecommand{\hypersetup}[1]{}

\hypersetup{plainpages=false}
\hypersetup{pdfpagemode=UseNone}
\hypersetup{bookmarksnumbered=true}
\hypersetup{pdfstartview=FitH}
\hypersetup{colorlinks=false}
\hypersetup{citebordercolor={.5 1 .5}}
\hypersetup{urlbordercolor={.5 1 1}}
\hypersetup{linkbordercolor={1 .7 .7}}

\providecommand{\arxivref}[2]{\href{http://arxiv.org/abs/#1}{#2}}
\providecommand{\doiref}[2]{\href{http://dx.doi.org/#1}{#2}}

\providecommand{\href}[2]{#2}
\providecommand{\arxivlink}[1]{\href{http://arxiv.org/abs/#1}{arxiv:#1}}

\newcommand{\eq}{\begin{equation}}
\newcommand{\eqx}{\end{equation}}
\newcommand{\eqs}{\begin{equation*}}
\newcommand{\eqsx}{\end{equation*}}
\newcommand{\eqn}{\begin{eqnarray}}
\newcommand{\eqnx}{\end{eqnarray}}
\newcommand{\alg}{\begin{align}}
\newcommand{\algx}{\end{align}}

\newcommand{\f}[2]{\frac{#1}{#2}}

\newcommand{\lm}{\lambda}

\renewcommand{\th}{\theta}
\newcommand{\sg}{\sigma}
\newcommand{\Sg}{\Sigma}
\newcommand{\dl}{\delta}
\newcommand{\Dl}{\Delta}

\newcommand{\bt}{\beta}
\newcommand{\om}{\omega}

\newcommand{\eps}{\varepsilon}
\newcommand{\qqqq}{\quad\quad\quad\quad}

\DeclareMathOperator{\arcsinh}{arcsinh}

\newcommand{\tr}{\mbox{\rm tr}\,}

\newcommand{\nn}{{\cal N}}

\newcommand{\oo}[1]{{\cal O}\left(#1\right)}

\newcommand{\sinpt}{\sin \f{p}{2}}
\newcommand{\qs}{qs}

\begin{document}


\thispagestyle{empty}
\phantomsection
\addcontentsline{toc}{section}{Title}

\begin{flushright}\footnotesize%
\texttt{\arxivlink{1012.3994}}\\
overview article: \texttt{\arxivlink{1012.3982}}%
\vspace{1em}%
\end{flushright}

\begingroup\parindent0pt
\begingroup\bfseries\ifarxiv\Large\else\LARGE\fi
\hypersetup{pdftitle={Review of AdS/CFT Integrability, Chapter III.5: L\"uscher corrections}}%
Review of AdS/CFT Integrability, Chapter III.5:\\
L\"uscher corrections
\par\endgroup
\vspace{1.5em}
\begingroup\ifarxiv\scshape\else\large\fi%
\hypersetup{pdfauthor={Romuald A. Janik}}%
Romuald A.\ Janik
\par\endgroup
\vspace{1em}
\begingroup\itshape
Institute of Physics\\
Jagiellonian University\\
ul.\ Reymonta 4\\ 
30-059 Krak\'ow\\
Poland
\par\endgroup
\vspace{1em}
\begingroup\ttfamily
romuald@th.if.uj.edu.pl
\par\endgroup
\vspace{1.0em}
\endgroup

\begin{center}
\includegraphics[width=5cm]{TitleIII5.mps}
\vspace{1.0em}
\end{center}

\paragraph{Abstract:}
In integrable quantum field theories the large volume spectrum is
given by the Bethe Ansatz. The leading corrections, due to virtual 
particles circulating around the cylinder, are encoded in so-called
L{\"u}scher corrections. In order to apply these techniques to the AdS/CFT correspondence one has to generalize these corrections to the case of generic dispersion relations and to multiparticle states. 
We review these various generalizations and the applications of L{\"u}scher's corrections to the study of the worldsheet QFT of the superstring in 
$AdS_5 \times S^5$ and, consequently, to anomalous dimensions of operators in 
$\nn=4$ SYM theory.

\ifarxiv\else
\fi

\ifarxiv\else
\fi
\hypersetup{pdfkeywords={AdS/CFT correspondence, integrability, integrable quantum field theory}}%

\newpage


\section{Introduction}

For many integrable systems the main question that one is interested in is the understanding of the energy spectrum for the system of a given size $L$. The size of the system in question may be discrete, like the number of sites of a spin chain or other kind of lattice system, or continuous, like the circumference of a cylinder on which a given integrable field theory is defined.

The first answer to this question for a wide variety of integrable systems is generically given in terms of Bethe equations. These are equations for a set of (complex) numbers $p_i$ of the form
\eq
\label{e.bethe}
1=e^{i p_j L} \prod_{k:k\neq j}^N S(p_j,p_k)
\eqx  
Once a solution $\{p_j\}_{j=1..N}$ is found, the energy is obtained through an additive formula
\eq
\label{e.enbethe}
E=\sum_{j=1}^N E(p_j)
\eqx
where $E(p)$ and $S(p,p')$ are (known) functions characterizing the given integrable system. In practice, for generic integrable systems, these equations become the more complicated nested Bethe equations, with a system of equations instead of (\ref{e.bethe}), with additional auxillary unknowns appearing in (\ref{e.bethe}) but not in the energy formula (\ref{e.enbethe}).
All this is described in detail in two other chapters of this review 
\cite{chapABA,chapSMat}.

Bethe equations of the type described above appear both in the case of discrete integrable spin chains and continuous two-dimensional integrable quantum field theories. Moreover they also appear as equations for the anomalous dimensions of single trace operators in the $\nn=4$ four-dimensional SYM theory and in various other contexts. 

Now comes the fundamental difference between the various classes of integrable systems. For integrable spin chains, like the Heisenberg XXX, XXZ etc. models, the Bethe ansatz equations are \emph{exact} and the energies given by 
(\ref{e.enbethe}) are the exact eigenvalues of the spin chain hamiltonian.
On the other hand, for two-dimensional integrable quantum field theories, the answer provided by (\ref{e.bethe}) and (\ref{e.enbethe}) is only valid for asymptotically large sizes of the cylinder $L$. There are corrections which arise due to the quantum field theoretical nature of the system, namely virtual particles circulating around the cylinder and their interaction with the physical particles forming a given energy state. For a single particle in a relativistic QFT, L\"uscher derived formulas \cite{Luscher:1985dn} for the leading corrections. The goal of this chapter is to review the subsequent generalizations and applications of L\"uscher corrections within the AdS/CFT correspondence. Let us note in passing that there may be also some intermediate cases like the Hubbard model as considered in \cite{Rej:2005qt}, where the situation is not so clear.

Once one goes beyond these leading corrections by say decreasing the size of the system, one has to include the effects of multiple virtual corrections which becomes quite complicated, and have never been attempted so far. Fortunately, for integrable quantum field theories, there exists a technique of Thermodynamic Bethe Ansatz --- TBA \cite{Zamolodchikov:1989cf} (and/or Noninear Integral Equations --- NLIE \cite{Destri:1994bv,Fioravanti:1996rz}), which allows for finding the exact energy spectrum and thus effectively resumming all these virtual corrections. This is, however, technically very involved, so even for the cases where it is known, L\"uscher corrections remain an efficient calculational tool. These exact treatments are described in the chapters \cite{chapTBA} and \cite{chapTrans} of this review.

As a final note, let us mention that for anomalous dimensions in the $\nn=4$ SYM theory, the Bethe equations break down due to so-called wrapping interactions. This will be discussed in more detail in section 3 (see also the chapter
\cite{chapHigher}). Since according to the AdS/CFT correspondence anomalous dimensions are exactly equal to the energies of string states in $AdS_5 \times S^5$, which are just the energy levels of the two-dimensional integrable worldsheet quantum field theory, this violation of Bethe ansatz equations is in fact quite natural and can be quantitatively described using the formalism of 
L\"uscher corrections for two dimensional QFT.

The plan of this chapter is as follows. First, after introducing L\"uscher's original formulas, we will describe the various derivations of (generalized versions of) these formulas -- a diagrammatic one, through a large volume expansion of TBA equations and through a Poisson resummation of quadratic fluctuations. Then we will review recent applications of generalized L\"uscher corrections within the context of the AdS/CFT correspondence.

\section{L{\"u}scher formulas}

L{\"u}scher derived universal formulas for the leading large $L$ mass shift (w.r.t. the particle mass in infinite volume) of a single particle state when the theory is put on a cylinder of size $L$. The universality means that the value of the leading correction is determined completely by the infinite volume S-matrix of the theory. This relation does not depend on integrability at all, and is even valid for arbitrary QFT's in higher number of dimensions, however it is most useful for two dimensional integrable field theories for which we very often know the exact analytical expression for the S-matrix.

\begin{figure}
\centerline{\includegraphics[height=6cm]{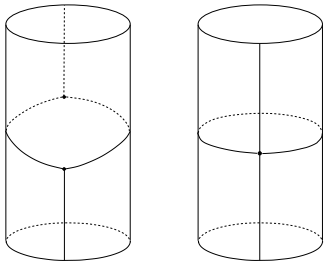}}
\caption{Spacetime interpretation of L{\"u}scher's formulas --- $\mu$-term (left) and F-term (right).}
\end{figure}

The leading mass correction is given as a sum of two terms -- the F-term
\eq
\label{e.fluscher}
\Delta m_F (L) = -m \int_{-\infty}^\infty
\f{d\th}{2\pi} e^{-m L \cosh \th} \cosh \th \sum_b
\left(S_{ab}^{ab}\left(\th+i \f{\pi}{2} \right)-1 \right)
\eqx
and the $\mu$-term
\eq
\label{e.muluscher}
\Delta m_\mu (L) =  -\f{\sqrt{3}}{2} m\sum_{b,c} M_{abc} (-i)
\,{\mbox{\rm res}}_{\th=2\pi i/3}\, S_{ab}^{ab}(\th) \cdot
e^{-\f{\sqrt{3}}{2}m L}
\eqx
quoted here, for simplicity, for a two dimensional theory with particles of the same mass \cite{Klassen:1990ub}. $S_{ab}^{ab}(\th)$ is the (infinite volume) $S$-matrix element, and $M_{abc}=1$ if $c$ is a bound state of $a$ and $b$ and zero
otherwise. These two terms have a distinct spacetime interpretation depicted in Figure~1. The F-term corresponds to the interaction of the physical particle with a virtual particle circulating around the cylinder, while the $\mu$-term corresponds to the splitting of the particle into two others which will then recombine after circulating around the cylinder.

In order to apply the above formulas to the case of the worldsheet QFT of the superstring in $AdS_5 \times S^5$ (in generalized light cone gauge -- see \cite{Arutyunov:2009ga} for a detailed review), one has to generalize L{\"u}scher's original formulas in two directions. 

Firstly, the worldsheet QFT is not relativistic. The dispersion relation for elementary excitations is
\eq
\label{e.adsdisp}
E(p)=\sqrt{1+16 g^2 \sin^2 \f{p}{2}}
\eqx
and moreover, there is no analog of a Lorentz symmetry, which brings about the fact that the S-matrix is a nontrivial function of two independent momenta instead of just the rapidity difference $\th\equiv \th_1-\th_2$ as in the case of relativistic theories.
Secondly, due to the level matching condition of the string, the physical states, corresponding to operators in $\nn=4$ SYM, have vanishing total momentum (or a multiple of $2\pi$). Since single particle states with $p=0$ are protected by supersymmetry, all states interesting from the point of view of gauge theory are neccessarily multiparticle states. 

Consequently, one has to generalize L{\"u}scher corrections to theories with quite generic dispersion relations and also to multiparticle states.

We will describe these generalizations at the same time showing how 
L{\"u}scher corrections can be derived in many different and apparently unrelated ways.

\subsection{Diagrammatic derivation}

The diagrammatic derivation was the original one used by L{\"u}scher in 
\cite{Luscher:1985dn}. Its advantage is that it is the most general, does not assume integrability and is even valid in any number of dimensions. Its drawback, however, is that it is very difficult to generalize to multiparticle states or higher orders.
On the other hand it is easy to extend to theories with generic dispersion relations which was done in \cite{Janik:2007wt}. We will present a sketch of this derivation here applicable to a theory with the dispertion relation
\eq
E^2=\eps^2(p)
\eqx
which encompasses both the relativistic dispersion relation $\eps^2(p)=m^2+p^2$, as well as the AdS one $\eps^2(p)=1+16g^2 \sin^2 p/2$.

The starting point is the observation that the dispersion relation is encoded, as the mass shell condition, in the pole structure of the Green's function. Hence to find the leading large $L$ corrections, one has to evaluate how the Green's function is modified at finite volume. It is convenient to translate the problem into a modification of the 1PI (1-particle irreducible) self energy defined by
\eq
\label{e.ginv}
G(p)^{-1} = \eps_E^2+\eps^2(p)-\Sg_L(p)
\eqx 
The renormalization scheme is fixed by requiring that the self energy and its first derivatives vanish on-shell (at infinite volume). The shift of the energy, following from (\ref{e.ginv}) becomes
\eq
\dl \eps_L = -\f{1}{2\eps(p)} \Sg_L(p)
\eqx

\begin{figure}[t]
\centerline{\includegraphics[width=13cm]{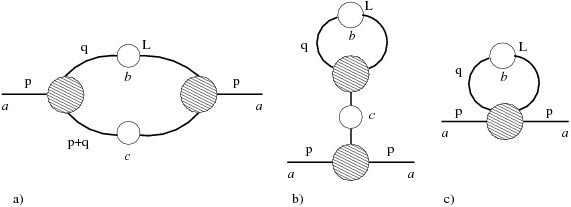}}
\caption{The graphs giving a leading finite size correction to the
  self energy: a) $I_{abc}$, b) $J_{abc}$, c) $K_{ab}$. The filled
  circles are the vertex functions $\Gamma$, empty circles represent the
  2-point Green's function. The letter $L$ represents the factor of
  $e^{-iq^1 L}$ and the letters in italics label the type of
  particles.}
\end{figure}

The propagator in a theory at fixed circumference can be obtained from the infinite volume one through averaging over translations $x \to x+n L$. In momentum space this will correspond to distributing factors of $e^{inp^1 L}$ over all lines.
In the next step, we assume, following \cite{Luscher:1985dn,Klassen:1990ub}, that the dominant corrections at large $L$ will be those graphs which have only a single such factor with $n=\pm 1$. Picking $n=-1$ for definiteness, any such graph belongs to one of the three classes shown in Figure~2. Thus
\eq
\Sg_L= \f{1}{2}\left( \sum_{bc} I_{abc} +\sum_{bc} J_{abc} +\sum_b
K_{ab} \right)
\eqx
Now, one has to shift the contour of integration over the loop \emph{spatial} 
momentum into the complex plane. Due to the exponent $e^{-ip^1 L}$, the integral over the shifted contour will be negligible and the main contribution will come from crossing a pole of a Green's function in one of the graphs of Figure~2. This is the crucial point for arriving at L{\"u}scher's corrections. Taking the residue effectively puts the line in question \emph{on-shell}, thus reducing the two dimensional loop integral to a single dimensional one. It is very convenient to eliminate the spatial momentum using the mass shell condition, and leave the last integral over Euclidean energy which we denote by $q$. The on-shell condition becomes
\eq
{q}^2+\eps(p^1)^2=0
\eqx
which in the case of the $AdS_5 \times S^5$ superstring theory leads to
\eq
p^1=-i 2 \arcsinh \f{\sqrt{1+q^2}}{4g}
\eqx
Plugging this back into the exponential factor $e^{-ip^1 L}$ leads to the term which governs the overall magnitude of the L{\"u}scher correction
\eq
e^{-ip^1 L}  = e^{-L \cdot E_{TBA}(q)}=e^{-L \cdot 2 \arcsinh
  \f{\sqrt{1+q^2}}{4g}} 
\eqx
We will analyze the physical meaning of this formula in section 3.

The mass shell condition has also another, equally important, consequence. Since the line is on-shell, in the integrand we may cut it open thus effectively transforming the graphs of the 2-point 1PI self energy into those of a 4-point forward Green's function. Keeping track of all the necessary factors gives
\eq
\label{e.gabab}
\Sg_L= \int \f{dq}{2\pi} \f{i}{\eps^2(p^1)'} \cdot e^{-L \cdot E_{TBA}(q)} 
\cdot \sum_b (-1)^{F_b} G_{abab}(-p,-q,p,q) 
\eqx
The $p$ appearing in the argument of $G_{abab}$ is the spatial momentum of the physical particle, while $q$ is the Euclidean energy of the virtual one.
In the final step, one links the 4-point forward Green's function with the forward S-matrix element arriving at L{\"u}scher's F-term formula generalized to a generic dispertion relation:
\eq
\label{e.fdiag}
\dl \eps^{F}_a = -\int_{-\infty}^\infty \f{dq}{2\pi}  \left(1-
\f{\eps'(p)}{\eps'(q)} \right) \cdot e^{-i\qs L} \cdot \sum_b (-1)^{F_b}
\left( S_{ba}^{ba}(q,p)-1 \right)
\eqx
with the same conventions for the arguments of $\eps'$ and $S_{ba}^{ba}$ as described below (\ref{e.gabab}).
The $\mu$-term arises in the process of shifting the contours by localizing on a residue of the above formula. It is thus given just by the residue of 
(\ref{e.fdiag}). For further details consult \cite{Janik:2007wt}. Let us mention that for relativistic theories one can perform a more detailed analysis concerning the contribution of $\mu$-terms \cite{Klassen:1990ub}. In particular, $\mu$-terms are expected to contribute only if, in the spacetime diagram shown in Figure~1, both particles move forward in time (i.e. have positive real part of the energy).
This analysis has not been done rigorously for the $AdS_5 \times S^5$ case.

The diagrammatic derivation presented above is very general and does not require integrability. Moreover the difference between a theory with diagonal and non-diagonal scattering is quite trivial. One can go from the simpler case of a single particle species to the most general case of nondiagonal scattering just by substituting the scalar S-matrix by an appropriate supertrace of the nondiagonal S-matrix. Generalizing this property to multiparticle states leads to a simple shortcut for obtaining multiparticle L{\"u}scher corrections --- one can first obtain the formulas for a simple theory with a single particle in the spectrum, and then generalize to the generic case by trading the product of the scalar S-matrices for a supertrace of the product of the nondiagonal ones. 
We will present this derivation in the following section.

\subsection{Multiparticle L{\"u}scher corrections from TBA}

In this section we will show how \emph{multiparticle} L{\"u}scher corrections arise from the Thermodynamic Bethe Ansatz. Here, we will be able to obtain these more powerful results using significantly stronger assumptions. In particular we will assume that the theory in question is integrable with diagonal scattering. Then, as explained above, we will use the expected very universal dependence of L{\"u}scher corrections on the S-matrix to conjecture the general versions valid for any integrable theory with a nondiagonal S-matrix (for which TBA equations are much more complicated).

As explained in \cite{chapTBA}, TBA equations are derived by trading the complicated problem of finding the (ground state) energy of the theory at finite volume for the much simpler one of computing a thermal partition function of the theory with space and time interchanged through a double Wick rotation. In the latter case, since one is dealing with the theory at almost infinite volume, Bethe ansatz is exact and can be used to evaluate the partition function. Hence the energies and momenta in the following are those of the spacetime interchanged one (\emph{aka} mirror theory) related to the energy and momentum of the original theory through
\eq
\tilde E =ip \qqqq  \tilde p=iE
\eqx 
The ground state TBA equation for the theory with a single type of particle takes the form
\eq
\label{e.tba}
\eps(z)= L\tilde E(z) + \int \f{dw}{2\pi i}\,
(\partial_w \log S(w,z)) \log\left(1+e^{-\eps(w)}\right)
\eqx
and the ground state energy is obtained from the solution $\eps(z)$ through
\eq
E= - \int \f{dz}{2\pi } \, \tilde p'(z) \log \left(1+e^{-\eps(z)}\right)
\eqx
In order to describe excited states, one uses an analytical continuation trick due to Dorey and Tateo \cite{Dorey:1996re} that essentially introduces additional source terms into 
(\ref{e.tba}). These source terms are generated by singularities of the integrand $1+e^{-\eps(z_i)}=0$, which, through integration by parts and an evaluation through residues give rise to additional source terms on the r.h.s. of (\ref{e.tba})
\eq
\label{e.tbagen}
\eps(z)=L\tilde E +\log S(z_1,z)+\log S(z_2,z)+ \int \f{dw}{2\pi i}\,
(\partial_w \log S(w,z)) \log\left(1+e^{-\eps(w)}\right) 
\eqx
and additional contributions to the energy
\eq
E=E(z_1)+E(z_2)- \int \f{dz}{2\pi} \, \tilde p'(z) \log
\left(1+e^{-\eps(z)}\right)
\eqx
where we quote the result with just two additional singularities.

It is quite nontrivial what kind of source terms to introduce for the theory at hand. If a theory does not have bound states and $\mu$-terms, the rule of thumb is that for each physical particle a single source term has to be included (this happens in the case of e.g. the sinh-Gordon model). On the other hand, for a theory with $\mu$-terms, like the SLYM, at least two source terms correspond to a single physical particle 
(see \cite{Dorey:1996re,Dorey:1997rb}).

Now in order to obtain the L{\"u}scher corrections, we have to perform a large volume expansion of these equations. Solving (\ref{e.tbagen}) by iteration, neglecting the integral term and inserting this approximation into the energy formula gives
\eqn
E &=& E(z_1)+E(z_2)- \int \f{dz}{2\pi} \, \tilde p' e^{-L \tilde E(z)}
\f{1}{S(z_1,z) S(z_2,z)} \nonumber\\
&=& E(z_1)+E(z_2)- \int \f{dq}{2\pi}  \, e^{-L \tilde E(q)} S(z,z_1)
S(z,z_2)
\eqnx
We recognize at once an integral of the F-term type (with $q \equiv \tilde p$) in addition to the sum of energies of the individual particles. There is a subtlety here, namely one has to dynamically impose the equations for the positions of the singularitites 
\eq
\label{e.quant}
1+e^{-\eps(z_i)}=0
\eqx 
If we insert here the same approximation as we have just used in the formula for the energy, we will recover the Bethe equations
\eq
e^{i L p_1}=S(p_1,p_2)
\eqx
However, in L{\"u}scher's corrections we should keep all leading exponential terms. Therefore for the quantization condition (\ref{e.quant}), we have to use also the first nontrivial iteration of (\ref{e.tbagen}). This will give rise to modifications of the Bethe quantization conditions.
The quantization conditions $\eps(z_i)=i\pi$ becomes
\eqn
0&=&\underbrace{\log\{ e^{iL p_1} S(z_2,z_1)\}}_{BY_1} + \underbrace{ \int
\f{dw}{2\pi i }\,( \partial_w S(w,z_1)) S(w,z_2) e^{-L \tilde E(w)}}_{\Phi_1}
\\ 
0&=&\underbrace{\log\{ e^{iL p_2} S(z_1,z_2)\}}_{BY_2} + \underbrace{ \int
\f{dw}{2\pi i}\, S(w,z_1) (\partial_w S(w,z_2)) e^{-L \tilde E(w)}}_{\Phi_2}
\eqnx
Since the integrals are exponentially small we may solve these equations in
terms of corrections to the Asymptotic Bethe Ansatz (ABA) giving
\eqn\label{BAmod.delatap1}
\f{\partial BY_1}{\partial p_1} \dl p_1 + \f{\partial BY_1}{\partial p_2} \dl
p_2 +\Phi_1 &=& 0 \\ \label{BAmod.delatap2}
\f{\partial BY_2}{\partial p_1} \dl p_1 + \f{\partial BY_2}{\partial p_2} \dl
p_2 +\Phi_2 &=& 0 
\eqnx
The final formula for the energy thus takes the form
\eq
\label{e.fin}
E=E(p_1)+E(p_2) + E'(p_1)\dl p_1+ E'(p_2)\dl p_2 - \int \f{dq}{2\pi}\, e^{-L
\tilde E} S(z,z_1) S(z,z_2)
\eqx
For nondiagonal scattering, we expect that the above formula will get modified just by exchanging the products of scalar S-matrices by a supertrace of a product of real \emph{matrix} S-matrices. This generalization has been proposed in 
\cite{Bajnok:2008bm}.
In the F-term integrand we will thus get the transfer matrix (c.f. \cite{chapTrans}) or more precisely its eigenvalue\footnote{We present below the case when the
physical particles forming the multiparticle state scatter between themselves
diagonally}
\eq
e^{i \delta(\tilde{p}\vert p_1,\dots ,p_N)} =(-1)^F\left[S_{a_{1}a}^{a_{2}a}(\tilde{p},p_{1})
S_{a_{2}a}^{a_{3}a}(\tilde{p},p_{2})\dots S_{a_{N}a}^{a_{1}a}(\tilde p,p_{N})\right]
\eqx
where we also substituted the complex rapidities used earlier by more physical
momenta.
The BY condition reads as 
\eq
2n_{k}\pi=BY_{k}(p_{1},\dots p_{n})+\delta\Phi_{k}=p_{k}L-i\log
\left[\prod_{k\neq j}S_{aa}^{aa}(p_{k},p_{j})\right]+\delta\Phi_{k}
\eqx
with the correction to these equations given by 
\eq
\delta\Phi_{k}=-\int_{-\infty}^{\infty}\frac{d\tilde{p}}{2\pi}
(-1)^{F}
\left[S_{a_{1}a}^{a_{2}a}(\tilde{p},p_{1})\dots\frac{\partial
S_{a_{k}a}^{a_{k+1}a}(\tilde{p},p_{k})}{\partial\tilde{p}} 
\dots S_{a_{N}a}^{a_{1}a}(\tilde p,p_{N})\right]
e^{-\tilde{\epsilon_{a_1}}(\tilde{p})L}
\eqx
The final correction then reads as 
\begin{align}
\label{e.fads}
E(L)&=\sum_{k}\epsilon(p_{k})-\sum_{j,k}\frac{d\epsilon(p_{k})}{dp_{k}} 
\left(\frac{\delta BY_{k}}{\delta p_{j}}\right)^{-1}\delta\Phi_{j} \nonumber\\ 
&-\int_{-\infty}^{\infty} \frac{d\tilde{p}}{2\pi} \sum_{a_1,\ldots,a_N}
(-1)^F  \left[ S_{a_{1}a}^{a_{2}a}(\tilde{p},p_{1})S_{a_{2}a}^{a_{3}a}
(\tilde{p},p_{2})\dots S_{a_{N}a}^{a_{1}a}(p,p_{N})\right]e^{-\tilde{\epsilon_{a_1}}(\tilde{p})L}
\end{align}
For theories with $\mu$-terms, we expect that the corresponding $\mu$-terms will be obtained by localizing the integrals on the poles of the S-matrix.

\subsection{Poisson resummation of fluctuations}

In this section we will present a simple, very physical, derivation of 
L{\"u}scher's F-term formula from a summation over quadratic fluctuations. 
Although this approach requires the most restrictive assumptions, it is quite intuitive and gives a new perspective on the origin of L{\"u}scher's corrections.

For simplicity we will just present the derivation for a particle with vanishing momentum, analogous to L{\"u}scher's original formulas. A more general case is treated\footnote{In ref. \cite{Heller:2008at}, a minus sign will have to be included in the second term in eq. (13) there.} in 
\cite{Heller:2008at}.

Consider a soliton at rest which is put on a very large cylinder, so large that we may neglect the effect of the deformation of the solution. Now a small fluctuation very far from the soliton will just be an excitation of the vacuum, so can be treated as another soliton\footnote{Here we use `soliton' as a generic term which includes e.g. `breathers' in the sine-Gordon model.} (more precisely a single particle state). This small `fluctuation' soliton will scatter on the stationary one and will get a phase shift expressible in terms of the S-matrix (which we assume here to be diagonal)
\eq
\label{e.phase}
S^{ba}_{ba}(k,p)=e^{i \dl_{ba}(k,p)}
\eqx
Due to the finite size of the cylinder, the momentum of the `fluctuation' soliton will have to be quantized giving
\eq
k_{n}=\frac{2\pi n}{L}+\frac{\delta_b(k_{n})}{L}
\eqx
We now have to perform a summation over the energies of the fluctuations
\eq
\label{e.enfluct}
\dl \eps_{naive} =\f{1}{2} \sum_b \sum_{n=-\infty}^\infty (-1)^{F_b}
\left(\eps(k_{n}) -\eps(k_{n}^{(0)})\right)
\eqx
where the energies of fluctuations around the vacuum 
(with $k_{n}^{(0)}=2\pi n/L$) have been subtracted out.

The key result of \cite{Heller:2008at} is that L{\"u}scher's corrections are exactly the leading exponential terms ($m=\pm 1$) in the Poisson resummation 
\eq
\sum_{n=-\infty}^\infty F\left(\f{2\pi n}{L} \right) = \f{L}{2\pi}
\sum_{m=-\infty}^\infty  \int_{-\infty}^{+\infty} F(t) e^{-im L t} dt
\eqx
of (\ref{e.enfluct}). The relevant terms will be
\eq
\label{e.start}
\dl \eps=\f{L}{4\pi} Re \int_{-\infty}^\infty e^{i L t}
(\epsilon(k(t)) -\epsilon(t)) dt
\eqx
where $k(t)=t+\dl(k(t))/L$ is the quantization condition, the solution of which we do not need explicitly. Now, after a sequence of integration by parts and a simple change of variables (see \cite{Heller:2008at} for details) we can rewrite (\ref{e.start}) as 
\eq
\dl \eps =\frac{1}{4 \pi i} Re \int_{-\infty}^{+\infty} e^{-i L k}
(e^{i\delta(k)} -1) \epsilon'(k) dk = \frac{1}{4 \pi i} Re \int_{-\infty}^{+\infty} e^{-i L k}
(S^{ba}_{ba}(k,p) -1) \epsilon'(k) dk
\eqx
which is essentially L{\"u}scher's F-term but evaluated on the physical line. We should now shift the contour to ensure that the exponent is strictly real and decreasing at infinity giving rise to L{\"u}scher's corrections. Here the  boundary terms require a case by case analysis. Also $\mu$-terms may be generated when in the process of shifting the contour we would encounter bound state poles. The above derivation shows that evaluating L{\"u}scher F-term contributions is equivalent  to computing directly 1-loop energy shifts around the corresponding classical solution. Although one has to be careful in this interpretation when one evaluates the phase shifts (\ref{e.phase}) exactly and not only semiclassically.

\section{Applications of generalized L{\"u}scher's corrections in the AdS/CFT correspondence}

In this section we will briefly review various applications of generalized L{\"u}scher's corrections in the context of the integrable worldsheet QFT of the superstring in $AdS_5 \times S^5$. Due to the AdS/CFT correspondence, the energy levels of this theory (energies of string states) are identified with the anomalous dimensions of the corresponding gauge theory operators. In this way, the intrinsically two-dimensional methods may be applied to the four-dimensional $\nn=4$ SYM theory.

Before we review the obtained results let us first discuss the generic magnitude of L{\"u}scher corrections in this context.

As we saw from the derivations, the order of magnitude of the F-term formula is essentially governed by the exponential factor \cite{Ambjorn:2005wa}
\eq
e^{-L \tilde{E}(\tilde{p})}
\eqx
where $\tilde{E}$ and $\tilde{p}\equiv q$ are the energy and momentum of a theory with a double Wick rotation exchanging space and time -- called `mirror theory' \cite{Arutyunov:2007tc}.
For the case at hand we have
\eq
e^{-L \cdot 2 \arcsinh \f{\sqrt{Q^2+q^2}}{4g}} 
\eqx
where $Q=1$ corresponds to the fundamental particle (magnon) and $Q>1$ labels the bound states of the theory.

In the strong coupling limit, this expression becomes
\eq
{e^{-Q\f{L}{2g}}}_{\big | Q=1} = e^{-\f{2\pi L}{\sqrt{\lm}}} 
\eqx
which is the typical finite size fluctuation effect observed for spinning strings \cite{SchaferNameki:2006gk,SchaferNameki:2006ey}. We also see that at strong coupling, the contribution of bound states is exponentially supressed, so one can just consider the fundamental magnons circulating around the cylinder.

The $\mu$-term, which arises from the F-term by taking residues also appears at strong coupling. It's magnitude at strong coupling for a single magnon can be estimated to be
\eq
\label{e.muestimate}
e^{-\f{2\pi J}{\sqrt{\lm} \sinpt}}
\eqx
where $p$ is the momentum of the physical magnon. We see that the exponential
term gives a stronger suppression than the F-term, however, the terms differ in
the scaling of the prefactor with the coupling. The F-term is associated with
quantum effects, while the $\mu$-term appears already in the classical contrbution
hence the F-term is supressed by a factor of $\sqrt{\lm}$ w.r.t. the $\mu$-term.
Let us note that the link between $\mu$-terms and classical solutions is stil
to a large extent not understood. We may get another qualitative estimate from
the formula (\ref{e.muestimate}) for classical finite-gap solutions which may be
considered to arise in the worldsheet theory as a state of very many particles,
each of which will presumably have a very small momentum. Then (\ref{e.muestimate})
suggests that the $\mu$-term should be completely negligible for such states.

At weak coupling, we obtain a quite different formula
\eq
\label{e.weakest}
\f{ \# \,g^{2L}}{(Q^2+q^2)^{L}}
\eqx
Firstly, we see that the effect of the virtual corrections only starts at a certain loop order, from the point of view of gauge theory perturbative expansion. Up to this order the Bethe equations are in fact exact. Such a behaviour is wholly due to the nonstandard AdS dispersion relation 
(\ref{e.adsdisp}). The loop order at which these corrections start to contribute is related to the size of the gauge theory operator in question. This is very good, as just at that order we expect a new class of Feynman graphs to appear in the perturbative computation. These are the so-called `wrapping corrections' and are given by graphs where, in the computation of a two point function relevant for extracting anomalous dimensions, at least one propagator crosses all vertical legs. From the very start 
\cite{Beisert:2004hm} (see also \cite{Sieg:2005kd}), these graphs were expected not to be described by the Asymptotic Bethe Ansatz. Their identification with (possibly multiple) L{\"u}scher corrections was first proposed in \cite{Ambjorn:2005wa}.
  
Secondly, we see that at weak coupling, all bound states contribute at the same order. This makes the computation of wrapping effects at weak coupling more complicated, but at the same time more interesting, as they are sensitive to much finer details of the worldsheet QFT than at strong coupling.

The corrections to L\"uscher formulas are very difficult to quantify. Even in the
relativistic case there are no formulas for the leading corrections.
These would be multiple wrapping graphs and hence a $0^{th}$ order estimate of
their relative magnitude would be another exponential term. At strong coupling we
would thus probably see a mixture of the first double wrapping graphs for magnons
with ordinary single wrapping graphs for the first $Q=2$ bound states. At weak coupling, the next wrapping correction would generically have a relative magnitude
of $g^{2L}$ although there might also be factors of $g$ coming from the prefactor
which we do not control so the loop order for subleading multiple wrapping 
corrections is not precisely determined. 

Let us finish this section with a brief note on the elusive nature of $\mu$-terms.
Physical arguments based on the relativistic spacetime picture of the $\mu$-term
diagram, amounting to the requirement that the produced virtual particles propagate forward in time suggests that at weak coupling $\mu$-terms should not appear since
the bound state is heavier than the fundamental magnon. Explicit computations for
the Konishi operator and twist-2 operators (see section 3.2 below) confirm this
intuition. Yet, at strong coupling the $\mu$-term definitely contributes to
the giant magnon finite size dispersion relation. It is still not understood
how and when does this occur, especially in terms of the proposed exact TBA
formulations.

\subsection{Strong coupling results}

An excitation of the worldsheet theory with momentum $p \sim \oo{1}$ has an energy which scales as $\sqrt{\lm}$ characteristic of a classical string solution. Such a solution has been found in \cite{Hofman:2006xt} and is called the `giant magnon'. Subsequently, corrections to its energy were found when the excitation was considered on a cylinder of finite size $J$. The resulting correction was evaluated from the deformed classical solution in \cite{Arutyunov:2006gs,Astolfi:2007uz} to be
\eq
\label{e.stringafz}
\dl E_{string}=-\f{\sqrt{\lm}}{\pi} \cdot \f{4}{e^2} \cdot \sin^3 \f{p}{2}
\cdot e^{-\f{2\pi J}{\sqrt{\lm} \sinpt}} \equiv
-g \cdot \f{16}{e^2} \cdot \sin^3 \f{p}{2} \cdot
e^{-\f{2}{g \sinpt} J}
\eqx 
In \cite{Janik:2007wt}, the above expression was recovered from L{\"u}scher's corrections. The exponential term is different from the one appearing in the F-term formula however it turns out that it is exactly the term appearing in the $\mu$-term, when we find the residue of the F-term expression at the bound state pole. 

The prefactor comes from evaluating the residue of the (super)trace of the forward S-matrix at the bound state pole. A very curious feature of the above expression is the contribution of the dressing factor, which, at strong coupling, has an expansion
(see \cite{chapSProp})
\eq
\sg^2 = \exp \Bigl( g \,\chi_{AFS}+\chi_{HL}+ \sum_{n=2}^\infty \f{1}{g^{n-1}} \chi_n  \Bigr)
\eqx
Naively, one may expect that only the first two terms would give a contribution, however it turns out that due to the special kinematics of the bound state pole, \emph{all} $\chi_{2n}$ contribute. The evaluation of this 
contribution is quite nontrivial with a divergent series appearing, which can be resummed using Borel resummation. The result exactly reproduces 
(\ref{e.stringafz}).

Among further developments, finite size contributions to dyonic giant magnons were analyzed \cite{Hatsuda:2008gd}, quantum fluctuations were linked with the F-term 
\cite{Gromov:2008ie,Heller:2008at}, similar computations were also done for giant magnons and dyonic magnons in the ABJM theory 
\cite{Grignani:2008te,Bombardelli:2008qd,Lukowski:2008eq,Ahn:2008wd,Ahn:2010eg}. In addition finite size corrections were evaluated for open strings (which corresponds to L{\"u}scher corrections in a boundary integrable field theory \cite{Bajnok:2004tq}) \cite{Correa:2009mz,Bajnok:2010ui}.

One can also analyze L{\"u}scher corrections for classical spinning strings. There the picture is quite different from the giant magnons. The spinning string solutions arise as a superposition of very many excitations, all with very small momenta. So the $\mu$-term exponential factor will be very much supressed and the dominant correction will arise from the F-term. The F-term integrand can be evaluated in terms of the transfer matrix directly in terms of the Bethe root distributions describing the spinning string in question. Alternatively, an analysis of these issues have been done from the algebraic curve perspective in \cite{Gromov:2008ec}.

\subsection{Weak coupling results}

L{\"u}scher's corrections are particularly interesting when applied in the weak coupling regime corresponding to perturbative gauge theory. There, they provide the only calculational method to compute wrapping corrections apart from a direct perturbative computation which usually is prohibitively complicated
(see \cite{chapHigher}). Calculations based on generalized L{\"u}scher's corrections are typically much simpler and allow to obtain 4- and 5- loop gauge theory results which cannot be obtained using other means.

From a more theoretical perspective, the agreement of L{\"u}scher corrections with perturbative gauge theory results is interesting as it gives a nontrivial quantitative test of the AdS/CFT correspondence, as well as of our understanding of the fine details of the worldsheet QFT of the $AdS_5 \times S^5$ superstring. Moreover, it is very interesting to realize that the breakdown of the Asymptotic Bethe Ansatz for anomalous dimensions in the four dimensional gauge theory occurs \emph{exactly} in a way characteristic of a \emph{two dimensional} quantum field theory (and thus characteristic of string theory).  

A natural testing ground for these methods is the Konishi operator 
$\tr \Phi_i^2$ (or equivalently $\tr XZXZ-\tr X^2 Z^2$, 
$\tr DZDZ-\tr Z D^2 Z$), which is the shortest operator not protected by supersymmetry. 

From the string perspective, it corresponds to a two particle state in the worldsheet QFT on a cylinder of size $J=2$. Despite the fact that $J$ is so small, we may expect to get an exact answer from L{\"u}scher corrections at least at 4- and 5- loop level due to the estimate (\ref{e.weakest}).
Since at weak coupling all bound states contribute at the same order, we have to perform a summation over all bound states and their polarization states and use the bound state-fundamental magnon S-matrix. There is a further subtlety here, which does not appear in relativistic systems. In the physical theory, the bound states discovered in \cite{Dorey:2006dq} are in the symmetric representation, while states in the antisymmetric representation are unstable. On the other hand, in the mirror theory, the physical bound states are in the antisymmetric representation \cite{Arutyunov:2007tc}, and in fact it is these antisymmetric bound states which have to be taken into account when computing 
L{\"u}scher's corrections.

\begin{figure}
\centerline{\includegraphics[height=4cm]{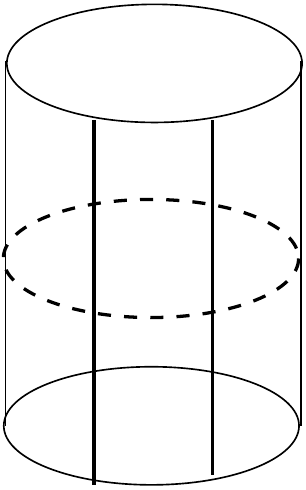}}
\caption{The single L{\"u}scher graph entering the computation of the four loop Konishi anomalous dimension.}
\end{figure}

Performing the computation yields the result for the 4-loop wrapping correction to the anomalous dimension of the Konishi operator \cite{Bajnok:2008bm}:
\eq
\Delta_{w}^{(8)} = 324 +864 \zeta(3) -1440 \zeta(5)
\eqx
which is in exact agreement with direct perturbative computations using both supergraph techniques \cite{Fiamberti:2007rj,Fiamberti:2008sh} and component Feynman graphs \cite{Velizhanin:2008jd}. The string computation is much simpler as it involves evaluating just the single graph shown in Figure~3.

In another development, wrapping corrections for twist two operators
\eq
\tr ZD^M Z+ \text{permutations}
\eqx
were computed. Here, the main motivation for performing this computation was the fact there are stringent analytical constraints on the structure of the anomalous dimensions $\Dl(M)$ as a function of $M$. In fact the disagreement, at 4 loops, between the behaviour of the Bethe Ansatz $\Dl(M)$ for $M=-1+\om$ and 
gauge theory constraints from the BFKL (Balitsky-Fadin-Kuraev-Lipatov) 
and NLO BFKL equations describing
high energy scattering in the Regge limit \cite{BFKL} were the first quantitative indication that the Asymptotic Bethe Ansatz breaks down 
\cite{Kotikov:2007cy}.

In \cite{Bajnok:2008qj}, the anomalous dimensions of twist two operators were evaluated at 4 loop level using L{\"u}scher corrections for an $M$-particle state. The wrapping correction was found to exactly compensate the mismatch between the Bethe Ansatz and BFKL expectations.

Subsequently, the leading wrapping corrections for the lowest lying twist-three operators were also determined from L{\"u}scher corrections 
\cite{Beccaria:2009eq}. These occur at 5 loop level. Another class of operators which was considered were single particle states 
\cite{Beccaria:2009hg} and the Konishi operator \cite{Ahn:2010yv} 
in the $\bt$ deformed theory. 
These results agree with direct field theoretical computations when avaiable 
\cite{Fiamberti:2008sn,Fiamberti:2009jw}.

In all the above computations of the leading wrapping corrections there were significant simplifications. Firstly, the wrapping modifications of the Bethe Ansatz quantization condition did not appear. Secondly, the dressing factor of the S-matrix also did not contribute.

Once one moves to subleading perturbative wrapping order (5-loop for Konishi and twist two, and 6-loop for twist three), both of these effects start to play a role. The modification of the Bethe Ansatz quantization is particularly interesting, as it is only in its derivation that the convolution terms in TBA equations contribute. In contrast to the simple single component TBA equation presented here, the structure of the TBA equations proposed for the $AdS_5 \times S^5$ system is very complicated \cite{Bombardelli:2009ns,Gromov:2009tv,Gromov:2009bc,Arutyunov:2009ur,Arutyunov:2009ux}. So L{\"u}scher corrections may be a nontrivial cross-check for these proposals.
In addition, due to the kinematics of the scattering between the physical particle and the mirror particle, it turns out that already at 5 loops, an infinite set of coefficients of the BES/BHL dressing phase contributes to the answer.

A key difficulty in performing such a computation is the possibility of testing the answer. Fortunately we have at our disposal two independent consistency checks. Firstly, at weak coupling we do not expect the appearance of $\mu$-terms which implies that a sum over residues of certain dynamical poles in the integrand has to cancel after summing over all bound states. Secondly, the transcendental structure of the final answer should be quite simple, while the subexpressions involve very complicated expressions which should cancel out in the final answer. In addition, for the case of twist two operators, one can use the numerous stringent constraints on the analytic structure comming from BFKL, NLO BFKL, reciprocity etc.

In \cite{Bajnok:2009vm}, the five loop wrapping correction to the Konishi anomalous dimension was derived
\eq
\Delta_w^{(10)} =-11340+2592\zeta(3) -5184\zeta(3)^2 -11520\zeta(5)
+30240\zeta(7)
\eqx
while in \cite{Lukowski:2009ce} a {\it tour-de-force} computation was performed for twist two operators at five loops. Subsequently twist three operators were also considered at subleading wrapping order in \cite{Velizhanin:2010cm}.

Recently, the five loop result coming from L{\"u}scher corrections was confirmed by expanding the exact TBA equations at large volume first numerically \cite{Arutyunov:2010gb}, and then analytically 
\cite{Balog:2010xa,Balog:2010vf}. Finally, subsubleading (6-loop) wrapping corrections were considered for single impurity operators in the $\bt$ deformed theory
\cite{Bajnok:2010ud}.

\section{Summary and outlook}

L{\"u}scher's corrections situate themselves in the middle ground between Bethe Ansatz and a full fledged solution of two dimensional integrable quantum field theories in the guise of Thermodynamic Bethe Ansatz or Nonlinear Integral Equations. They encode effects of an explicitly quantum field theoretical nature, namely virtual corrections associated with the topology of a cylinder. In this way L{\"u}scher's corrections may be seen to differentiate between spin chain like systems, where the Bethe Ansatz is exact and quantum field theories, for which the Bethe Ansatz is only a large volume approximation.

In this review, we have presented various ways of arriving at L{\"u}scher's corrections, some of them more or less rigorous, others more conjectural. The fact that the methods are quite different one from the other serves as an important cross check of these results. It would be, however, quite interesting to extend some of these methods in various directions e.g. the diagrammatic calculations to multiparticle states and subleading wrapping. Recently, the multiparticle L{\"u}scher corrections proposed in \cite{Bajnok:2008bm} were tested in \cite{Gromov:2008gj,Balog:2009ze,Arutyunov:2010gb,Balog:2010xa,Balog:2010vf}. It would be interesting to obtain some kind of universal understanding how the structure necessary for L{\"u}scher corrections is encoded in the very complicated nondiagonal TBA systems.

With respect to the concrete applications of L{\"u}scher corrections in the AdS/CFT correspondence there are still some loose ends like the rather mysterious formula for the finite size corrections of the giant magnon in the $\bt$ deformed theory \cite{Bykov:2008bj}. Apart from that, the agreement between the computations based on L{\"u}scher corrections, which typically involve a single graph, and the very complicated four loop gauge theory computations involving hundreds or even many thousands of graphs suggests that there is some very nontrivial hidden structure in the perturbative expansion. It would be very interesting to understand whether it could be understood in any explicit way.

\bigskip

\noindent{}{\bf Acknowledgments:} This work was supported by Polish science funds as a research project N N202 105136 (2009-2011).

\phantomsection
\addcontentsline{toc}{section}{\refname}

\begin{thebibliography}{10}
\ifx\href\asklfhas\newcommand{\href}[2]{#2}\fi
\ifx\arxivref\asklfhas\newcommand{\arxivref}[2]{\href{http://arxiv.org/abs/#1}%
{#2}}\fi
\ifx\doiref\asklfhas\newcommand{\doiref}[2]{\href{http://dx.doi.org/#1}{#2}}\fi
\raggedright
\small
\parskip 0pt

\bibitem{chapABA}
M.~Staudacher,
\textit{``Review of AdS/CFT Integrability, Chapter III.1: Bethe Ans\"atze and
  the R-Matrix Formalism''},
\texttt{\arxivref{1012.3990}{arxiv:1012.3990}}.

\bibitem{chapSMat}
C.~Ahn and R.~Nepomechie,
\textit{``Review of AdS/CFT Integrability, Chapter III.2: Exact world-sheet
  S-matrix''},
\texttt{\arxivref{1012.3991}{arxiv:1012.3991}}.

\bibitem{Luscher:1985dn}
M.~Luscher,
\textit{``{Volume Dependence of the Energy Spectrum in Massive Quantum Field
  Theories. 1. Stable Particle States}''},
\textsf{\doiref{10.1007/BF01211589}{Commun.~Math.~Phys.~104,~177~(1986)}}.

\bibitem{Rej:2005qt}
A.~Rej, D.~Serban and M.~Staudacher,
\textit{``{Planar $\mathcal{N}$ = 4 gauge theory and the Hubbard model}''},
\textsf{\doiref{10.1088/1126-6708/2006/03/018}{JHEP~0603,~018~(2006)}},
\texttt{\arxivref{hep-th/0512077}{hep-th/0512077}}.

\bibitem{Zamolodchikov:1989cf}
A.~B.~Zamolodchikov,
\textit{``{THERMODYNAMIC BETHE ANSATZ IN RELATIVISTIC MODELS. SCALING THREE
  STATE POTTS AND LEE-YANG MODELS}''},
\textsf{\doiref{10.1016/0550-3213(90)90333-9}{Nucl.~Phys.~B342,~695~(1990)}}.

\bibitem{Destri:1994bv}
C.~Destri and H.~J.~De~Vega,
\textit{``{Unified approach to thermodynamic Bethe Ansatz and finite size
  corrections for lattice models and field theories}''},
\textsf{\doiref{10.1016/0550-3213(94)00547-R}{Nucl.~Phys.~B438,~413~(1995)}},
\texttt{\arxivref{hep-th/9407117}{hep-th/9407117}}.

\bibitem{Fioravanti:1996rz}
D.~Fioravanti, A.~Mariottini, E.~Quattrini and F.~Ravanini,
\textit{``{Excited state Destri-De Vega equation for sine-Gordon and restricted
  sine-Gordon models}''},
\textsf{\doiref{10.1016/S0370-2693(96)01409-8}{Phys.~Lett.~B390,~243~(1997)}},
\texttt{\arxivref{hep-th/9608091}{hep-th/9608091}}.

\bibitem{chapTBA}
Z.~Bajnok,
\textit{``Review of AdS/CFT Integrability, Chapter III.6: Thermodynamic Bethe
  Ansatz''},
\texttt{\arxivref{1012.3995}{arxiv:1012.3995}}.


\bibitem{chapTrans}
V.~Kazakov and N.~Gromov,
\textit{``Review of AdS/CFT Integrability, Chapter III.7: Hirota Dynamics for
  Quantum Integrability''},
\texttt{\arxivref{1012.3996}{arxiv:1012.3996}}.

\bibitem{chapHigher}
C.~Sieg,
\textit{``Review of AdS/CFT Integrability, Chapter I.2: The spectrum from
  perturbative gauge theory''},
\texttt{\arxivref{1012.3984}{arxiv:1012.3984}}.

\bibitem{Klassen:1990ub}
T.~R.~Klassen and E.~Melzer,
\textit{``{On the relation between scattering amplitudes and finite size mass
  corrections in QFT}''},
\textsf{\doiref{10.1016/0550-3213(91)90566-G}{Nucl.~Phys.~B362,~329~(1991)}}.

\bibitem{Arutyunov:2009ga}
G.~Arutyunov and S.~Frolov,
\textit{``{Foundations of the AdS$_5$ $\times$ S$^5$ Superstring. Part I}''},
\textsf{\doiref{10.1088/1751-8113/42/25/254003}{J.~Phys.~A42,~254003~(2009)}},
\texttt{\arxivref{0901.4937}{arxiv:0901.4937}}.

\bibitem{Janik:2007wt}
R.~A.~Janik and T.~Lukowski,
\textit{``{Wrapping interactions at strong coupling -- the giant magnon}''},
\textsf{\doiref{10.1103/PhysRevD.76.126008}{Phys.~Rev.~D76,~126008~(2007)}},
\texttt{\arxivref{0708.2208}{arxiv:0708.2208}}.

\bibitem{Dorey:1996re}
P.~Dorey and R.~Tateo,
\textit{``{Excited states by analytic continuation of TBA equations}''},
\textsf{\doiref{10.1016/S0550-3213(96)00516-0}{Nucl.~Phys.~B482,~639~(1996)}},
\texttt{\arxivref{hep-th/9607167}{hep-th/9607167}}.

\bibitem{Dorey:1997rb}
P.~Dorey and R.~Tateo,
\textit{``{Excited states in some simple perturbed conformal field
  theories}''},
\textsf{\doiref{10.1016/S0550-3213(97)00838-9}{Nucl.~Phys.~B515,~575~(1998)}},
\texttt{\arxivref{hep-th/9706140}{hep-th/9706140}}.

\bibitem{Heller:2008at}
M.~P.~Heller, R.~A.~Janik and T.~Lukowski,
\textit{``{A new derivation of L\"{u}scher F-term and fluctuations around the
  giant magnon}''},
\textsf{\doiref{10.1088/1126-6708/2008/06/036}{JHEP~0806,~036~(2008)}},
\texttt{\arxivref{0801.4463}{arxiv:0801.4463}}.

\bibitem{Ambjorn:2005wa}
J.~Ambjorn, R.~A.~Janik and C.~Kristjansen,
\textit{``{Wrapping interactions and a new source of corrections to the
  spin-chain / string duality}''},
\textsf{\doiref{10.1016/j.nuclphysb.2005.12.007}{Nucl.~Phys.~B736,~288~(2006)}%
},
\texttt{\arxivref{hep-th/0510171}{hep-th/0510171}}.

\bibitem{Arutyunov:2007tc}
G.~Arutyunov and S.~Frolov,
\textit{``{On String S-matrix, Bound States and TBA}''},
\textsf{\doiref{10.1088/1126-6708/2007/12/024}{JHEP~0712,~024~(2007)}},
\texttt{\arxivref{0710.1568}{arxiv:0710.1568}}.

\bibitem{SchaferNameki:2006gk}
S.~Schafer-Nameki,
\textit{``{Exact expressions for quantum corrections to spinning strings}''},
\textsf{\doiref{10.1016/j.physletb.2006.03.033}{Phys.~Lett.~B639,~571~(2006)}},
\texttt{\arxivref{hep-th/0602214}{hep-th/0602214}}.

\bibitem{SchaferNameki:2006ey}
S.~Schafer-Nameki, M.~Zamaklar and K.~Zarembo,
\textit{``{How accurate is the quantum string Bethe ansatz?}''},
\textsf{\doiref{10.1088/1126-6708/2006/12/020}{JHEP~0612,~020~(2006)}},
\texttt{\arxivref{hep-th/0610250}{hep-th/0610250}}.

\bibitem{Beisert:2004hm}
N.~Beisert, V.~Dippel and M.~Staudacher,
\textit{``{A novel long range spin chain and planar $\mathcal{N}$ = 4 super
  Yang-Mills}''},
\textsf{\doiref{10.1088/1126-6708/2004/07/075}{JHEP~0407,~075~(2004)}},
\texttt{\arxivref{hep-th/0405001}{hep-th/0405001}}.

\bibitem{Sieg:2005kd}
C.~Sieg and A.~Torrielli,
\textit{``{Wrapping interactions and the genus expansion of the 2-point
  function of composite operators}''},
\textsf{\doiref{10.1016/j.nuclphysb.2005.06.011}{Nucl.~Phys.~B723,~3~(2005)}},
\texttt{\arxivref{hep-th/0505071}{hep-th/0505071}}.

\bibitem{Hofman:2006xt}
D.~M.~Hofman and J.~M.~Maldacena,
\textit{``{Giant magnons}''},
\textsf{\doiref{10.1088/0305-4470/39/41/S17}{J.~Phys.~A39,~13095~(2006)}},
\texttt{\arxivref{hep-th/0604135}{hep-th/0604135}}.

\bibitem{Arutyunov:2006gs}
G.~Arutyunov, S.~Frolov and M.~Zamaklar,
\textit{``{Finite-size effects from giant magnons}''},
\textsf{\doiref{10.1016/j.nuclphysb.2006.12.026}{Nucl.~Phys.~B778,~1~(2007)}},
\texttt{\arxivref{hep-th/0606126}{hep-th/0606126}}.

\bibitem{Astolfi:2007uz}
D.~Astolfi, V.~Forini, G.~Grignani and G.~W.~Semenoff,
\textit{``Gauge invariant finite size spectrum of the giant magnon,''},
\textsf{Phys.\ Lett.\  B{651} 329 (2007)},
\texttt{\arxivref{hep-th/0702043}{hep-th/0702043}}

\bibitem{chapSProp}
Vieira, P. and Volin, D.,
\textit{``Review of AdS/CFT Integrability, Chapter III.3: The dressing factor''},
\texttt{\arxivref{1012.3992}{arxiv:1012.3992}}.


\bibitem{Hatsuda:2008gd}
Y.~Hatsuda and R.~Suzuki,
\textit{``{Finite-Size Effects for Dyonic Giant Magnons}''},
\textsf{\doiref{10.1016/j.nuclphysb.2008.04.007}{Nucl.~Phys.~B800,~349~(2008)}%
},
\texttt{\arxivref{0801.0747}{arxiv:0801.0747}}.

\bibitem{Gromov:2008ie}
N.~Gromov, S.~Schafer-Nameki and P.~Vieira,
\textit{``{Quantum Wrapped Giant Magnon}''},
\textsf{\doiref{10.1103/PhysRevD.78.026006}{Phys.~Rev.~D78,~026006~(2008)}},
\texttt{\arxivref{0801.3671}{arxiv:0801.3671}}.

\bibitem{Grignani:2008te}
G.~Grignani, T.~Harmark, M.~Orselli and G.~W.~Semenoff,
\textit{``Finite size Giant Magnons in the string dual of N=6 superconformal
Chern-Simons theory''},
\textsf{JHEP 0812, 008 (2008)},
\texttt{\arxivref{arXiv:0807.0205}{arXiv:0807.0205}}.


\bibitem{Bombardelli:2008qd}
D.~Bombardelli and D.~Fioravanti,
\textit{``{Finite-Size Corrections of the CP$^3$ Giant Magnons: the L\"{u}scher
  terms}''},
\textsf{\doiref{10.1088/1126-6708/2009/07/034}{JHEP~0907,~034~(2009)}},
\texttt{\arxivref{0810.0704}{arxiv:0810.0704}}.

\bibitem{Lukowski:2008eq}
T.~Lukowski and O.~O.~Sax,
\textit{``{Finite size giant magnons in the SU(2) $\times$ SU(2) sector of
  AdS$_4$ $\times$ CP$^3$}''},
\textsf{\doiref{10.1088/1126-6708/2008/12/073}{JHEP~0812,~073~(2008)}},
\texttt{\arxivref{0810.1246}{arxiv:0810.1246}}.

\bibitem{Ahn:2008wd}
C.~Ahn and P.~Bozhilov,
\textit{``{Finite-size Effect of the Dyonic Giant Magnons in $\mathcal{N}$ = 6
  super Chern-Simons Theory}''},
\textsf{\doiref{10.1103/PhysRevD.79.046008}{Phys.~Rev.~D79,~046008~(2009)}},
\texttt{\arxivref{0810.2079}{arxiv:0810.2079}}.

\bibitem{Ahn:2010eg}
C.~Ahn, M.~Kim and B.-H.~Lee,
\textit{``{Quantum finite-size effects for dyonic magnons in the $AdS_4 x
  CP^3$}''},
\textsf{\doiref{10.1007/JHEP09(2010)062}{JHEP~1009,~062~(2010)}},
\texttt{\arxivref{1007.1598}{arxiv:1007.1598}}.

\bibitem{Bajnok:2004tq}
Z.~Bajnok, L.~Palla and G.~Takacs,
\textit{``{Finite size effects in quantum field theories with boundary from
  scattering data}''},
\textsf{\doiref{10.1016/j.nuclphysb.2005.03.021}{Nucl.~Phys.~B716,~519~(2005)}%
},
\texttt{\arxivref{hep-th/0412192}{hep-th/0412192}}.

\bibitem{Correa:2009mz}
D.~H.~Correa and C.~A.~S.~Young,
\textit{``{Finite size corrections for open strings/open chains in planar
  AdS/CFT}''},
\textsf{\doiref{10.1088/1126-6708/2009/08/097}{JHEP~0908,~097~(2009)}},
\texttt{\arxivref{0905.1700}{arxiv:0905.1700}}.

\bibitem{Bajnok:2010ui}
Z.~Bajnok and L.~Palla,
\textit{``{Boundary finite size corrections for multiparticle states and planar
  AdS/CFT}''},
\texttt{\arxivref{1010.5617}{arxiv:1010.5617}}.

\bibitem{Gromov:2008ec}
N.~Gromov, S.~Schafer-Nameki and P.~Vieira,
\textit{``{Efficient precision quantization in AdS/CFT}''},
\textsf{\doiref{10.1088/1126-6708/2008/12/013}{JHEP~0812,~013~(2008)}},
\texttt{\arxivref{0807.4752}{arxiv:0807.4752}}.

\bibitem{Dorey:2006dq}
N.~Dorey,
\textit{``{Magnon bound states and the AdS/CFT correspondence}''},
\textsf{\doiref{10.1088/0305-4470/39/41/S18}{J.~Phys.~A39,~13119~(2006)}},
\texttt{\arxivref{hep-th/0604175}{hep-th/0604175}}.

\bibitem{Bajnok:2008bm}
Z.~Bajnok and R.~A.~Janik,
\textit{``{Four-loop perturbative Konishi from strings and finite size effects
  for multiparticle states}''},
\textsf{\doiref{10.1016/j.nuclphysb.2008.08.020}{Nucl.~Phys.~B807,~625~(2009)}%
},
\texttt{\arxivref{0807.0399}{arxiv:0807.0399}}.

\bibitem{Fiamberti:2007rj}
F.~Fiamberti, A.~Santambrogio, C.~Sieg and D.~Zanon,
\textit{``{Wrapping at four loops in $\mathcal{N}$ = 4 SYM}''},
\textsf{\doiref{10.1016/j.physletb.2008.06.061}{Phys.~Lett.~B666,~100~(2008)}},
\texttt{\arxivref{0712.3522}{arxiv:0712.3522}}.

\bibitem{Fiamberti:2008sh}
F.~Fiamberti, A.~Santambrogio, C.~Sieg and D.~Zanon,
\textit{``{Anomalous dimension with wrapping at four loops in $\mathcal{N}$ = 4
  SYM}''},
\textsf{\doiref{10.1016/j.nuclphysb.2008.07.014}{Nucl.~Phys.~B805,~231~(2008)}%
},
\texttt{\arxivref{0806.2095}{arxiv:0806.2095}}.

\bibitem{Velizhanin:2008jd}
V.~N.~Velizhanin,
\textit{``{The four-loop anomalous dimension of the Konishi operator in N=4
  supersymmetric Yang-Mills theory}''},
\textsf{\doiref{10.1134/S0021364009010020}{JETP~Lett.~89,~6~(2009)}},
\texttt{\arxivref{0808.3832}{arxiv:0808.3832}}.

\bibitem{BFKL}
  L.~N.~Lipatov,
  { ``Reggeization of the vector meson and the
  vacuum singularity in nonabelian gauge theories,''}
  Sov.\ J.\ Nucl.\ Phys.\  {\bf 23} (1976) 338
  [Yad.\ Fiz.\  {\bf 23} (1976) 642];\\
  E.~A.~Kuraev, L.~N.~Lipatov and V.~S.~Fadin,
  { ``The Pomeranchuk singularity in nonabelian gauge theories,''}
  Sov.\ Phys.\ JETP {\bf 45} (1977) 199
  [Zh.\ Eksp.\ Teor.\ Fiz.\  {\bf 72} (1977) 377];\\
  I.~I.~Balitsky and L.~N.~Lipatov,
  { ``The Pomeranchuk singularity in Quantum Chromodynamics,''}
  Sov.\ J.\ Nucl.\ Phys.\  {\bf 28} (1978) 822
  [Yad.\ Fiz.\  {\bf 28} (1978) 1597].


\bibitem{Kotikov:2007cy}
A.~V.~Kotikov, L.~N.~Lipatov, A.~Rej, M.~Staudacher and V.~N.~Velizhanin,
\textit{``{Dressing and Wrapping}''},
\textsf{\doiref{10.1088/1742-5468/2007/10/P10003}{J.~Stat.~Mech.~0710,~P10003~%
(2007)}},
\texttt{\arxivref{0704.3586}{arxiv:0704.3586}}.

\bibitem{Bajnok:2008qj}
Z.~Bajnok, R.~A.~Janik and T.~Lukowski,
\textit{``{Four loop twist two, BFKL, wrapping and strings}''},
\textsf{\doiref{10.1016/j.nuclphysb.2009.02.005}{Nucl.~Phys.~B816,~376~(2009)}%
},
\texttt{\arxivref{0811.4448}{arxiv:0811.4448}}.

\bibitem{Beccaria:2009eq}
M.~Beccaria, V.~Forini, T.~Lukowski and S.~Zieme,
\textit{``{Twist-three at five loops, Bethe Ansatz and wrapping}''},
\textsf{\doiref{10.1088/1126-6708/2009/03/129}{JHEP~0903,~129~(2009)}},
\texttt{\arxivref{0901.4864}{arxiv:0901.4864}}.

\bibitem{Beccaria:2009hg}
M.~Beccaria and G.~F.~De~Angelis,
\textit{``{On the wrapping correction to single magnon energy in twisted N=4
  SYM}''},
\textsf{\doiref{10.1142/S0217751X09047375}{Int.~J.~Mod.~Phys.~A24,~5803~(2009)%
}},
\texttt{\arxivref{0903.0778}{arxiv:0903.0778}}.

\bibitem{Ahn:2010yv}
C.~Ahn, Z.~Bajnok, D.~Bombardelli and R.~I.~Nepomechie,
\textit{``{Finite-size effect for four-loop Konishi of the beta- deformed N=4
  SYM}''},
\textsf{\doiref{10.1016/j.physletb.2010.08.056}{Phys.~Lett.~B693,~380~(2010)}},
\texttt{\arxivref{1006.2209}{arxiv:1006.2209}}.

\bibitem{Fiamberti:2008sn}
F.~Fiamberti, A.~Santambrogio, C.~Sieg and D.~Zanon,
\textit{``{Single impurity operators at critical wrapping order in the
  beta-deformed $\mathcal{N}$ = 4 SYM}''},
\textsf{\doiref{10.1088/1126-6708/2009/08/034}{JHEP~0908,~034~(2009)}},
\texttt{\arxivref{0811.4594}{arxiv:0811.4594}}.

\bibitem{Fiamberti:2009jw}
F.~Fiamberti, A.~Santambrogio and C.~Sieg,
\textit{``{Five-loop anomalous dimension at critical wrapping order in N=4
  SYM}''},
\textsf{\doiref{10.1007/JHEP03(2010)103}{JHEP~1003,~103~(2010)}},
\texttt{\arxivref{0908.0234}{arxiv:0908.0234}}.

\bibitem{Bombardelli:2009ns}
D.~Bombardelli, D.~Fioravanti and R.~Tateo,
\textit{``{Thermodynamic Bethe Ansatz for planar AdS/CFT: a proposal}''},
\textsf{\doiref{10.1088/1751-8113/42/37/375401}{J.~Phys.~A42,~375401~(2009)}},
\texttt{\arxivref{0902.3930}{arxiv:0902.3930}}.

\bibitem{Gromov:2009tv}
N.~Gromov, V.~Kazakov and P.~Vieira,
\textit{``{Exact Spectrum of Anomalous Dimensions of Planar N=4 Supersymmetric
  Yang-Mills Theory}''},
\textsf{\doiref{10.1103/PhysRevLett.103.131601}{Phys.~Rev.~Lett.~103,~131601~(%
2009)}},
\texttt{\arxivref{0901.3753}{arxiv:0901.3753}}.

\bibitem{Gromov:2009bc}
N.~Gromov, V.~Kazakov, A.~Kozak and P.~Vieira,
\textit{``{Exact Spectrum of Anomalous Dimensions of Planar N = 4
  Supersymmetric Yang-Mills Theory: TBA and excited states}''},
\textsf{\doiref{10.1007/s11005-010-0374-8}{Lett.~Math.~Phys.~91,~265~(2010)}},
\texttt{\arxivref{0902.4458}{arxiv:0902.4458}}.

\bibitem{Arutyunov:2009ur}
G.~Arutyunov and S.~Frolov,
\textit{``{Thermodynamic Bethe Ansatz for the AdS$_5$ $\times$ S$^5$ Mirror
  Model}''},
\textsf{\doiref{10.1088/1126-6708/2009/05/068}{JHEP~0905,~068~(2009)}},
\texttt{\arxivref{0903.0141}{arxiv:0903.0141}}.

\bibitem{Arutyunov:2009ux}
G.~Arutyunov and S.~Frolov,
\textit{``{Simplified TBA equations of the $AdS_5 x S^5$ mirror model}''},
\textsf{\doiref{10.1088/1126-6708/2009/11/019}{JHEP~0911,~019~(2009)}},
\texttt{\arxivref{0907.2647}{arxiv:0907.2647}}.

\bibitem{Bajnok:2009vm}
Z.~Bajnok, A.~Hegedus, R.~A.~Janik and T.~Lukowski,
\textit{``{Five loop Konishi from AdS/CFT}''},
\textsf{\doiref{10.1016/j.nuclphysb.2009.10.015}{Nucl.~Phys.~B827,~426~(2010)}%
},
\texttt{\arxivref{0906.4062}{arxiv:0906.4062}}.

\bibitem{Lukowski:2009ce}
T.~Lukowski, A.~Rej and V.~N.~Velizhanin,
\textit{``{Five-Loop Anomalous Dimension of Twist-Two Operators}''},
\textsf{\doiref{10.1016/j.nuclphysb.2010.01.008}{Nucl.~Phys.~B831,~105~(2010)}%
},
\texttt{\arxivref{0912.1624}{arxiv:0912.1624}}.

\bibitem{Velizhanin:2010cm}
V.~N.~Velizhanin,
\textit{``{Six-Loop Anomalous Dimension of Twist-Three Operators in N=4
  SYM}''},
\textsf{\doiref{10.1007/JHEP11(2010)129}{JHEP~1011,~129~(2010)}},
\texttt{\arxivref{1003.4717}{arxiv:1003.4717}}.

\bibitem{Arutyunov:2010gb}
G.~Arutyunov, S.~Frolov and R.~Suzuki,
\textit{``{Five-loop Konishi from the Mirror TBA}''},
\textsf{\doiref{10.1007/JHEP04(2010)069}{JHEP~1004,~069~(2010)}},
\texttt{\arxivref{1002.1711}{arxiv:1002.1711}}.

\bibitem{Balog:2010xa}
J.~Balog and A.~Hegedus,
\textit{``{5-loop Konishi from linearized TBA and the XXX magnet}''},
\textsf{\doiref{10.1007/JHEP06(2010)080}{JHEP~1006,~080~(2010)}},
\texttt{\arxivref{1002.4142}{arxiv:1002.4142}}.

\bibitem{Balog:2010vf}
J.~Balog and A.~Hegedus,
\textit{``{The Bajnok-Janik formula and wrapping corrections}''},
\textsf{\doiref{10.1007/JHEP09(2010)107}{JHEP~1009,~107~(2010)}},
\texttt{\arxivref{1003.4303}{arxiv:1003.4303}}.

\bibitem{Bajnok:2010ud}
Z.~Bajnok and O.~e.~Deeb,
\textit{``{6-loop anomalous dimension of a single impurity operator from
  AdS/CFT and multiple zeta values}''},
\texttt{\arxivref{1010.5606}{arxiv:1010.5606}}.

\bibitem{Gromov:2008gj}
N.~Gromov, V.~Kazakov and P.~Vieira,
\textit{``{Finite Volume Spectrum of 2D Field Theories from Hirota
  Dynamics}''},
\textsf{\doiref{10.1088/1126-6708/2009/12/060}{JHEP~0912,~060~(2009)}},
\texttt{\arxivref{0812.5091}{arxiv:0812.5091}}.

\bibitem{Balog:2009ze}
J.~Balog and A.~Hegedus,
\textit{``{The finite size spectrum of the 2-dimensional O(3) nonlinear
  sigma-model}''},
\texttt{\arxivref{0907.1759}{arxiv:0907.1759}}.

\bibitem{Bykov:2008bj}
D.~V.~Bykov and S.~Frolov,
\textit{``{Giant magnons in TsT-transformed AdS$_5$ $\times$ S$^5$}''},
\textsf{\doiref{10.1088/1126-6708/2008/07/071}{JHEP~0807,~071~(2008)}},
\texttt{\arxivref{0805.1070}{arxiv:0805.1070}}.

\end{thebibliography}

\end{document}